\documentclass[fleqn,twoside]{article}
\usepackage{espcrc2}
\usepackage{psfig}
\usepackage{bm}

\begin{document}

\title{
{\vspace{-1.2em} \parbox{\hsize}{\hbox to \hsize 
{\hss  \normalsize IFUP-TH 2003/29, UPRF-2003-20}}} \\
The two-dimensional Wess-Zumino model 
in the Hamiltonian lattice formulation}
\author{
Matteo Beccaria\address{Dipartimento di Fisica
dell'Universit\`a di Lecce and I.N.F.N., Sezione di Lecce},
Massimo Campostrini\address{Dipartimento di Fisica ``Enrico Fermi''
dell'Universit\`a di Pisa and I.N.F.N., Sezione di Pisa}%
\thanks{Presented by M.\ Campostrini} and
Alessandra Feo\address{School of Mathematics, Trinity College,
Dublin 2, Ireland, \\ Dipartimento di Fisica dell'Universit\`a di Parma,
and I.N.F.N., Sezione di Parma}
}

\begin{abstract}
We investigate a Hamiltonian lattice version of the two-dimensional
Wess-Zumino model, with special emphasis to the pattern of
supersymmetry breaking.  Results are obtained by Quantum Monte Carlo
simulations and Density Matrix Renormalization Group techniques.
\end{abstract}

\maketitle

\section{INTRODUCTION AND PRESENTATION OF THE MODEL}

Most studies of lattice field theory are performed in the well-known
{\em Lagrangian\/} formalism, discretizing both space and time.  We
feel that it is important not to neglect the {\em Hamiltonian\/}
formalism \cite{KS}, which affords, e.g., a more immediate approach
to the mass spectrum and to the implementation of fermions; numerical
methods are quite different in the Lagrangian and in Hamiltonian
formalism, and comparison between the two can provide important test
of the methods as well as of universality.

We chose to study an Hamiltonian lattice version of the
two-dimensional Wess-Zumino model, described in Ref.~\cite{Trento};
the model enjoys an exact 1-dimensional supersymmetry algebra $H=Q^2$,
which suffices to preserve a few key properties of the continuum model
(which has a full 2-dimensional supersymmetry): the energy spectrum is
non-negative; states of nonzero energy appear in boson-fermion
doublets; spontaneous supersymmetry breaking is equivalent to a
strictly positive ground-state energy $E_0$.

The model is parametrized by the {\em prepotential\/}, an arbitrary
polynomial in the bosonic field $V(\phi)$; the model is
superrenormalizable, and the only renormalization needed is the normal
ordering of $V(\phi)$.

It is interesting to notice that the strong-coupling expansion
predicts supersymmetry breaking if and only if the degree of the
prepotential $V$ is even, while perturbation theory predicts
supersymmetry breaking if and only if $V$ has no zeroes; the two
contitions are quite different in the case, e.g., of quadratic $V$,
and it is interesting to study the crossover from strong to weak
coupling.

\section{NUMERICAL METHODS}

A standard numerical simulation technique that can be applied to the
model we are interested in is the Green Function Monte Carlo (GFMC)
algorithm~\cite{QMC}.  The interested reader may find a description of
GFMC applied to the present problem in Ref.~\cite{Trento}; we only
remark here that the GFMC algorithm generates a stochastic
representation of the ground-state wavefunction, from which
expectation values of observables can be computed; the variance of
observables can be kept under control by the introduction of a
population of $K$ {\em walkers\/} (field configurations), and an
extrapolation to $K\to\infty$ of the result is required.

Another numerical approach which has been applied with success to
1-dimensional Hamiltonian models is the Density-Matrix Renormalization
Group (DMRG)~\cite{DMRG,DMRGrev}; the parameters of the algorithm are the
dimension $S$ to which the local (1-site) Hilbert space is
truncated, and the number $M$ of eigenstates of the density matrix of
the full system that are retained; extrapolation to $M\to\infty$ and
$S\to\infty$ is needed.  DMRG is a deterministic algorithm and does
not suffer from the sign problem; it is possible to obtain directly
infinite-volume results.

\section{NUMERICAL RESULTS}

The most interesting property of the model is the pattern of
supersymmetry breaking, especially in the case of quadratic
prepotential; we study the case of even prepotential $V(\phi) =
\lambda_2\phi^2 + \lambda_0$, for which the strong-coupling expansion
predicts supersymmetry breaking for all values of $\lambda_0$, while
perturbation theory predicts unbroken supersymmetry for $\lambda_0<0$.

The model enjoys an approximate $\phi_m\to-\phi_m$, $n_m\to1-n_m$
symmetry (``parity''); supersymmetry breaking is incompatible with
parity breaking~\cite{Witten82}, and it is reasonable to expect only
two phases: unbroken supersymmetry and broken parity for
$\lambda_0<\lambda_c$, broken supersymmetry and unbroken parity for
$\lambda_0>\lambda_c$.

An analysis of the approach to the continuum limit, along a trajectory
of ``constant physics'' in the phase of broken supersymmetry was
presented in Ref.\ \cite{lattice2002}.  In the present paper, 
we study the model varying $\lambda_0$ at fixed $\lambda_2=0.5$.

\begin{figure}[tb]
\null\vskip2mm
\centerline{\psfig{figure=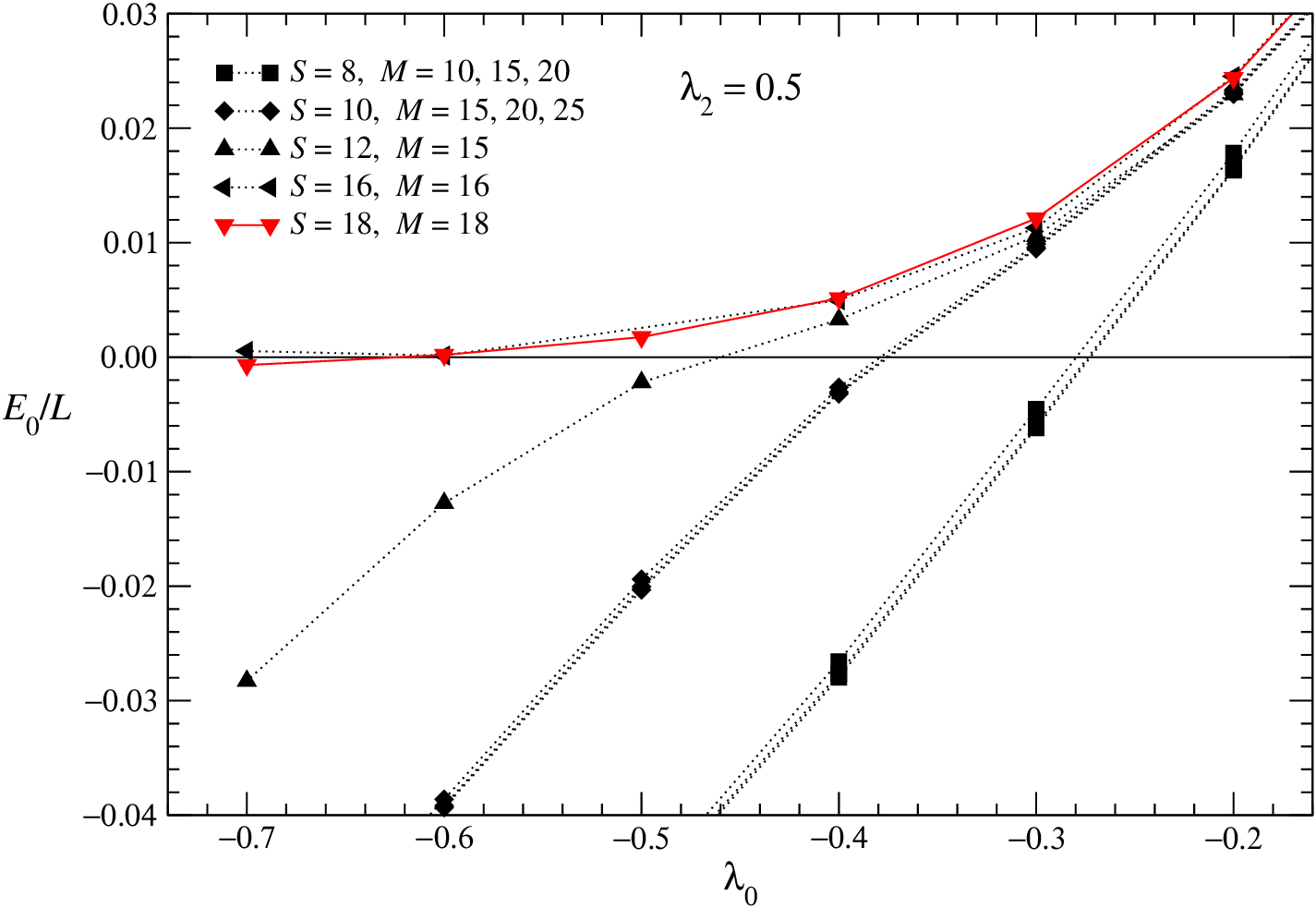,width=75mm}}
\vskip-8mm
\caption{Ground-state energy density vs.\ $\lambda_0$ obtained by DMRG
with different values of $S$ and $M$.}
\label{fig:DMRG_E0}
\end{figure}

\begin{figure}[tb]
\null\vskip2mm
\centerline{\psfig{figure=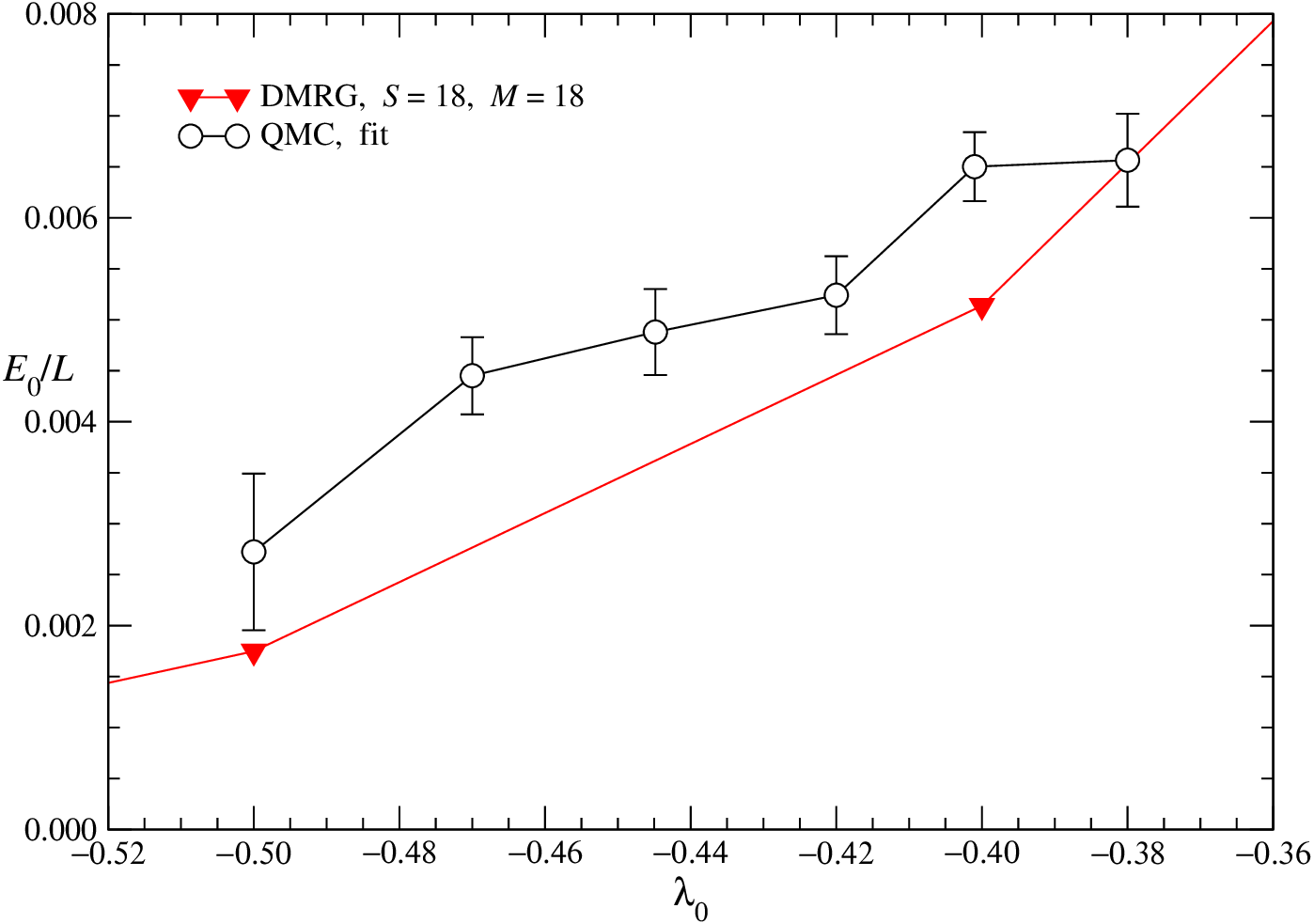,width=75mm}}
\vskip-8mm
\caption{Ground-state energy density vs.\ $\lambda_0$ obtained by DMRG
with different values of $S$ and $M$.}
\label{fig:GFMC_E0}
\end{figure}

We monitor supersymmetry breaking by looking at the ground-state
energy density $E_0/L$.  Infinite-volume DMRG results are shown in
Fig.~\ref{fig:DMRG_E0}; we remind that they must be extrapolated to
$S\to\infty$ and $M\to\infty$.
GFMC results are extrapolated to $L\to\infty$ and
$K\to\infty$ by a best fit to the form
\[ 
{E_0\over L} = {\cal E} + {a\over L} + {b\over L^2}
+ {c\over K^\nu} + {d\over L K^\nu},
\]
which gives a good $\chi^2$ (at least for $\lambda_0>-0.46$); the fit
gives $\nu\simeq0.5$, signaling a slow convergence to $K\to\infty$ (to
stabilize the fit, we ran with $K$ up to 5000 for our smallest lattice
$L=22$).  ${\cal E}$ is plotted in Fig.~\ref{fig:GFMC_E0}, together
with DMRG results for the highest values of $S$ and $M$; the agreement
is remarkable.

\begin{figure}[tb]
\null\vskip2mm
\centerline{\psfig{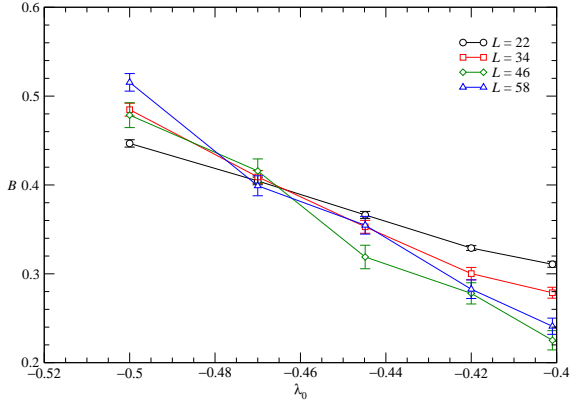}}
\vskip-8mm
\caption{The Binder cumulant $B$ for $K=200$.}
\label{fig:B_K200}
\end{figure}

\begin{figure}[tb]
\null\vskip2mm
\centerline{\psfig{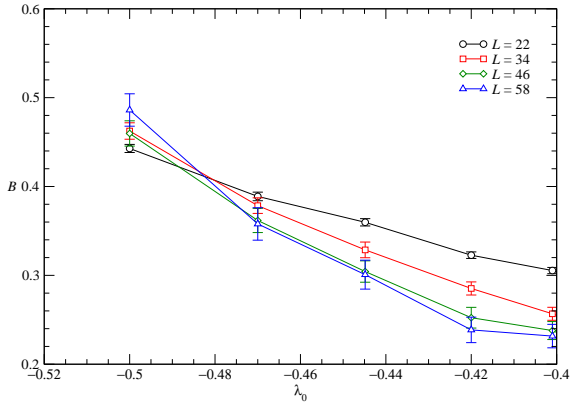}}
\vskip-8mm
\caption{The Binder cumulant $B$ for $K=500$.}
\label{fig:B_K500}
\end{figure}

\begin{figure}[tb]
\null\vskip2mm
\centerline{\psfig{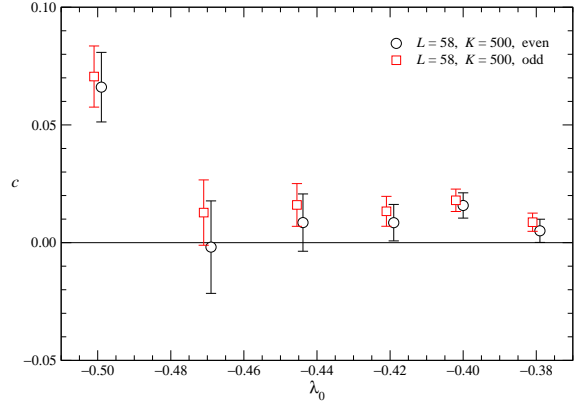}}
\vskip-8mm
\caption{The long-distance correlation of $\phi$ vs.\ $\lambda_0$ for
$L=58$ and $K=500$.}
\label{fig:c_K500}
\end{figure}

We study parity symmetry breaking looking at the Binder cumulant
\[
B = {\langle M^4\rangle\over\langle M^2\rangle^2}, \qquad
M = \sum_n \phi_n,
\]
where the sum excludes sites closer to the border than (typically) 6;
a good estimate of the transition point is the intersection of the
curves $B$ vs.\ $\lambda_0$ obtained at different values of $L$.  GFMC
results are shown in Figs.~\ref{fig:B_K200} and \ref{fig:B_K500}; they
are quite sensitive to $K$ and not very precise.
Therefore, we follow a different strategy and consider the connected
correlation function $G_d=\langle\phi_n\phi_{m}\rangle_c$ averaged
over all $n,m$ pairs with $|m-n|=d$, excluding pairs for which $m$ or
$n$ is closer to the border than (typically) 6; since we are using
staggered fermions, even and odd $d$ may correspond to different
physical channels.  The correlation function is quite different in the
two phases: for $\lambda_0\gg\lambda_c$, there is a marked difference
between even and odd distances, and $G_d$ is short-ranged; for
$\lambda_0\ll\lambda_c$, even and odd distances are equivalent, and
$G_d$ appears to be long-ranged.  We fit $G_d$ to the form $\exp(a+b
d) + c$, separately for even and odd $d$; the best fits give a good
$\chi^2$ if we remove the smallest distances, typically $d\le3$ for
the odd channel and $d\le6$ for the even channel.  The difference
between the two phases is apparent, e.g., in the plot of $c$ vs.\ 
$\lambda_0$, presented in Fig.~\ref{fig:c_K500}.


\begin{thebibliography}{9}

\bibitem{KS} J.~Kogut, L.~I.~Susskind, 
Phys.\ Rev.\ D11 (1975) 395; 
J.~Kogut, Rev.\ Mod.\ Phys.\ 51 (1979) 659.

\bibitem{Trento} 
M.~Beccaria, M.~Campostrini, A.~Feo, e-print {\tt hep-lat/0109005}, in
{\em Quantum Monte Carlo}, M.~Campostrini, M.~P.~Lombardo, and
F.~Pederiva eds., ETS, Pisa, 2001.

\bibitem{QMC} W.~von der Linden,
Phys. Rept.\ 220 (1992) 53.

\bibitem{DMRG}
C.~Zhang, E.~Jeckelmann, S.~R.~White,
Phys.\ Rev.\ Lett.\ 80 (1998) 2661;

\bibitem{DMRGrev}
K.~Hallberg, e-print {\tt cond-mat/0303557}, to appear in {\em
Theoretical Methods for Strongly Correlated Electrons}, D.~Senechal,
A.-M.\ Tremblay and C.~Bourbonnais eds., CRM Series in Mathematical
Physics, Springer, New York, 2003.


\bibitem{Witten82} E.~Witten, Nucl.\ Phys.\ B202 (1982) 253.

\bibitem{lattice2002} M.~Beccaria, M.~Campostrini, A.~Feo, 
Nucl.\ Phys.\ B (Proc.\ Suppl.) 119 (2002) 891.

\end{thebibliography}
\end{document}